# Annular Newtonian Poiseuille flow with pressure-dependent wall slip

Kostas D. Housiadas*
*Department of Mathematics,*
*University of the Aegean, Karlovassi, Samos 83200, Greece*
E-mail: housiada@aegean.gr
*Corresponding author

Evgenios Gryparis
*Department of Mathematics and Statistics,*
*University of Cyprus, PO Box 20537, 1678, Nicosia, Cyprus*
E-mail: gryparis.evgenios@ucy.ac.cy

Georgios C. Georgiou
*Department of Mathematics and Statistics,*
*University of Cyprus, PO Box 20537, 1678, Nicosia, Cyprus*
E-mail: georgios@ucy.ac.cy

**Abstract**

We investigate the effect of pressure-dependent wall slip on the steady Newtonian annular Poiseuille flow employing Navier's slip law with a slip parameter that varies exponentially with pressure. The dimensionless governing equations and accompanying auxiliary conditions are solved analytically up to second order by implementing a regular perturbation scheme in terms of the small dimensionless pressure-dependence slip parameter. An explicit formula for the average pressure drop, required to maintain a constant volumetric flowrate, is also derived. This is suitably post-processed by applying a convergence acceleration technique to increase the accuracy of the original perturbation series. The effects of pressure-dependent wall slip are more pronounced when wall slip is weak. However, as the slip coefficient increases, these effects are moderated and eventually eliminated as the perfect slip case is approached. The results show that the average pressure drop remains practically constant until the Reynolds number becomes sufficiently large. It is worth noting that all phenomena associated with pressure-dependent wall slip are amplified as the annular gap is reduced.

## 1. Introduction

The effect of wall slip in Poiseuille flows has been broadly discussed in many studies during the past decades. Nowadays, it is well-known that fluids such as polymer solutions, emulsions, yield stress fluids and suspensions, exhibit slip at solid surfaces [1-3]. Nevertheless, this behavior has also been reported for Newtonian liquids. Neto *et al.* [4] reviewed the phenomenon of wall slip of Newtonian liquids at solid interfaces via experiments, where they focused on the importance of surface roughness, wettability, and gaseous film or nanobubbles at interface. Such studies are considered significant in the fields of microfluidic and microelectromechanical devices. In addition, wall slip effects become more pronounced in the case of non-Newtonian fluids where leading to interesting phenomena and instabilities [2,5,6].

Both static and dynamic slip models have been reviewed by Hatzikiriakos [3] who highlighted their importance on the rheology and flow simulations of molten polymers. In static slip models, the slip velocity $u_w^*$, defined as the relative velocity of the adjacent fluid particles with respect to that of the wall, is assumed to be a function of the instantaneous value of the wall shear stress $\tau_w^*$, $u_w^* = u_w^*(\tau_w^*)$. A popular slip equation is the power-law one,

$$u_w^* = \beta^* \tau_w^{*m} \qquad (1)$$

where $\beta^*$ is the slip coefficient, which depends on temperature, normal stress and pressure, molecular parameters and characteristics of the fluid-wall interface, and $m$ is the power-law exponent [1,7,8,9]. When $\beta^* = 0$ the standard no-slip boundary condition is recovered, while when $m = 1$, Eq. (1) is simplified to the well-known Navier slip law [10]. Note that in the present work star superscripts denote dimensional quantities.

Experimental studies indicate that normal stresses which include pressure, as well as the past states of the local wall shear, can affect the slip velocity (see [3] and references therein). Though, the dependence of slip velocity on normal stresses is generally weaker compared to that on the wall shear stress [3,8]. Overall, the slip coefficient and the slip velocity are decreasing functions of the normal stresses which implies that wall slip in Poiseuille flow is weak or negligible upstream and becomes more important downstream near the tube exit [11]. Hill *et al.* [8] proposed the following slip law in which the slip coefficient decays exponentially with pressure $p^*$:

$$u_w^* = A^* \exp\left(-\varepsilon^*(p^* - p_{ref}^*)\right)\tau_w^* \qquad (2)$$

where $A^*$ is the slip coefficient at the reference pressure $p_{ref}^*$, and $\varepsilon^*$ is the pressure-dependence slip parameter. Slip laws similar to Eq. (2) were also employed by Hatzikiriakos & Dealy [9]. In addition, Tang & Kalyon [12-14] suggested a mathematical model describing the time-dependent pressure-induced flow of compressible polymer fluids subject to pressure-dependent slip Stewart [15] also developed a density-dependent slip law. Reviews of theoretical and experimental works on flows involving pressure-dependent wall slip can be found in [11,16].

Panaseti *et al.* [11] and Housiadas [17] employed the slip law of Eq. (2) and applied regular perturbation methods in order to derive analytical solutions for Poiseuille flows in channels and pipes of Newtonian and viscoelastic fluids, respectively. In both works the pressure decay parameter $\varepsilon^*$ served as the small parameter of the perturbation. Pressure-dependent wall slip has also been considered in other numerical and theoretical studies on Newtonian and non-Newtonian fluids [18,19].

In this paper we provide an asymptotic solution for concentric annular Newtonian Poiseuille flow with pressure-dependent wall slip, extending thus the work of Panaseti *et al.* [11]. Annular Poiseuille flows are encountered in a plethora of industrial applications such as in heat exchangers, biomedical devices and others. The effects of pressure-independent wall slip in this geometry has been investigated in the literature for both Newtonian [20,21] and non-Newtonian fluids [22,23].

The rest of the paper is organized as follows. The governing equations accompanied with the suitable auxiliary (boundary, symmetry, and integral) conditions are presented in Section 2, first in dimensional and then in non-dimensional form. A new alternative perturbation procedure is described in Section 3, where the analytical solution is derived up to second order in terms of the dimensionless pressure-dependent slip parameter. Unlike other problems solved utilizing regular perturbation schemes [11,24,25,26], in the case of annular tubes, the direct implementation of the asymptotic method to the governing equations in differential form, results in complex and very long expressions for the higher-order solutions. To prevent this, we converted the conservation laws into non-linear integrodifferential form, before applying the perturbation method. Section 4 discusses the most important results and effects of the pressure-dependent slip in terms of the Reynolds number, slip coefficient, and radii ratio. The main conclusions are summarized in Section 5.

**2. Governing equations**

We consider the isothermal, steady, incompressible, pressure-driven, axisymmetric flow of a Newtonian fluid between coaxial cylinders of length $L^*$ and inner and outer radii $\kappa R^*$ and $R^*$, respectively, where $0 < \kappa < 1$. We utilize the cylindrical coordinate system with the axis origin placed at the axis of symmetry at the entry plane of the annulus as illustrated in Fig.1. Due to axisymmetry, the velocity field is of the form $\mathbf{u}^* = u_z^* \mathbf{e}_z + u_r^* \mathbf{e}_r$. The flow is governed by the continuity equation

$$\frac{\partial u_r^*}{\partial r^*} + \frac{u_r^*}{r^*} + \frac{\partial u_z^*}{\partial z^*} = 0 \qquad (3)$$

and the *z*- and *r*-components of the momentum equation,

$$\rho^*\left(u_r^*\frac{\partial u_r^*}{\partial r^*}+u_z^*\frac{\partial u_r^*}{\partial z^*}\right)=-\frac{\partial p^*}{\partial r^*}+\eta^*\left[\frac{\partial}{\partial r^*}\left(\frac{1}{r^*}\frac{\partial}{\partial r^*}\left(r^*u_r^*\right)\right)+\frac{\partial^2 u_r^*}{\partial z^{*2}}\right] \quad (4)$$

and

$$\rho^*\left(u_r^*\frac{\partial u_z^*}{\partial r^*}+u_z^*\frac{\partial u_z^*}{\partial z^*}\right)=-\frac{\partial p^*}{\partial z^*}+\eta^*\left[\frac{1}{r^*}\frac{\partial}{\partial r^*}\left(r^*\frac{\partial u_z^*}{\partial r^*}\right)+\frac{\partial^2 u_z^*}{\partial z^{*2}}\right] \quad (5)$$

where $p^*$ is the pressure, $\rho^*$ is the constant mass density, and $\eta^*$ is the constant viscosity of the fluid.

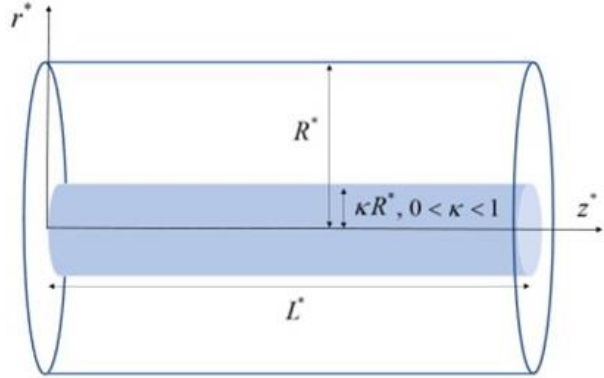

**Figure 1**: Geometry and symbols in annular Poiseuille flow.

We assume that pressure-dependent slip occurs along both the inner and the outer cylinders, while the normal to the walls velocity of the fluid vanishes (no penetration). Hence, the boundary conditions along the outer wall are

$$u_z^*+A^*\exp\left(-\varepsilon^*(p^*-p_{ref}^*)\right)\eta^*\left(\frac{\partial u_z^*}{\partial r^*}+\frac{\partial u_r^*}{\partial z^*}\right)=0 \quad \text{at } r^*=R^*,\ 0\leq z^*\leq L^* \quad (6)$$

$$u_r^*=0 \quad \text{at } r^*=R^*,\ 0\leq z^*\leq L^* \quad (7)$$

Along the inner cylinder the boundary conditions read:

$$u_z^*+A^*\exp\left(-\varepsilon^*(p^*-p_{ref}^*)\right)\eta^*\left(\frac{\partial u_z^*}{\partial r^*}+\frac{\partial u_r^*}{\partial z^*}\right)=0 \quad \text{at } r^*=\kappa R^*,\ 0\leq z^*\leq L^* \quad (8)$$

$$u_r^*=0 \quad \text{at } r^*=\kappa R^*,\ 0\leq z^*\leq L^* \quad (9)$$

Finally, we assume that the volumetric flow rate $\dot{Q}^*$ at the inlet cross-section is given,

$$2\pi\int_{\kappa R^*}^{R^*}u_z^*(r^*,0)r^*\,dr^*=\dot{Q}^* \quad (10)$$

and that the exit pressure at the outer cylinder is equal to a reference pressure:

$$p_{ref}^*=p^*(R^*,L^*) \quad (11)$$

## 2.1. Non-dimensionalization

To non-dimensionalize the governing equations using the mean velocity $u_m^*$ we employ the following characteristic velocity

$$u_c^* \equiv \frac{\dot{Q}^*}{2\pi R^{*2}} = \frac{1}{2}(1-\kappa^2)u_m^* \tag{12}$$

The dimensionless variables are then defined as follows:

$$U = \frac{u_z^*}{u_c^*}, \quad V = \frac{u_r^* L^*}{u_c^* R^*}, \quad P = \frac{(p^* - p_{ref}^*)R^{*2}}{\eta^* L^* u_c^*}, \quad r = \frac{r^*}{R^*}, \quad z = \frac{z^*}{L^*} \tag{13}$$

With these characteristic scales, the flow domain is $\Omega = \{(r,z) \mid \kappa \leq r \leq 1, 0 \leq z \leq 1\}$ and the dimensionless forms of Eqs. (3)-(5) become

$$\frac{\partial}{\partial r}(rV) + \frac{\partial}{\partial z}(rU) = 0 \tag{14}$$

$$a\,\mathrm{Re}\left(V\frac{\partial V}{\partial r} + U\frac{\partial V}{\partial z}\right) = -\frac{\partial P}{\partial r} + a\frac{\partial}{\partial r}\left(\frac{1}{r}\frac{\partial}{\partial r}(rV)\right) + a^2\frac{\partial^2 V}{\partial z^2} \tag{15}$$

and

$$\mathrm{Re}\left(V\frac{\partial U}{\partial r} + U\frac{\partial U}{\partial z}\right) = -\frac{\partial P}{\partial z} + \frac{1}{r}\frac{\partial}{\partial r}\left(r\frac{\partial U}{\partial r}\right) + a\frac{\partial^2 U}{\partial z^2} \tag{16}$$

where $a$ is the square of the aspect ratio of the tube and $\mathrm{Re}$ is the Reynolds number,

$$a \equiv \frac{R^{*2}}{L^{*2}}, \quad \mathrm{Re} \equiv \frac{\rho^* u_c^* R^{*2}}{\eta^* L^*} \tag{17}$$

The dimensionless auxiliary conditions closing the problem, valid for $0 < z < 1$, are:

$$U(1,z) = -A\exp(-\varepsilon P(1,z))\left.\frac{\partial U}{\partial r}\right|_{r=1} \tag{18}$$

$$U(\kappa,z) = A\exp(-\varepsilon P(\kappa,z))\left.\frac{\partial U}{\partial r}\right|_{r=\kappa} \tag{19}$$

$$V(\kappa,z) = V(1,z) = 0 \tag{20}$$

$$\int_\kappa^1 U(r,z)r\,dr = 1 \tag{21}$$

$$P(1,1) = 0 \tag{22}$$

where

$$A \equiv \frac{A^* \eta^*}{R^*} \tag{23}$$

is the dimensionless slip number, and

$$\varepsilon \equiv \frac{\eta^* L^* u^*_{z_C} \varepsilon^*}{R^{*2}} \tag{24}$$

is the dimensionless pressure-dependence slip decay coefficient, which serves as the small perturbation parameter in Section 3. It should be noted that when $A = 0$, the no-slip boundary condition is recovered and thus both slip velocities become zero. When $\varepsilon = 0$, wall slip is pressure-independent and thus the classic Navier slip law applies, $V = 0$ and the axial velocity is independent of the axial coordinate z, $U = U(r)$ [20,22].

## 3. Solution

Integrating Eq. (14) with respect to *r* and applying the no-penetration condition at the inner wall (Eq. (20)) gives:

$$V(r,z) = -\frac{1}{r}\frac{\partial}{\partial z}\int_{\kappa}^{r} U(y,z)y\,dy \tag{25}$$

Using the boundary condition (18) one obtains from the *z*-momentum Eq. (16) (see the Appendix) the following integrodifferential equation:

$$U(r,z) = U(\kappa,z)\left[1 + \frac{\kappa}{A}\exp(\varepsilon P(\kappa,z))\ln\left(\frac{r}{\kappa}\right)\right] + \mathrm{Re}\int_{\kappa}^{r} U(y,z)V(y,z)\,dy$$
$$-\frac{\partial}{\partial z}\left[\int_{\kappa}^{r}\left(\mathrm{Re}\,U^2(y,z) + p(y,z)\right)\ln\left(\frac{y}{r}\right)y\,dy\right] + a\frac{\partial^2}{\partial z^2}\left[\int_{\kappa}^{r} U(y,z)\ln\left(\frac{y}{r}\right)y\,dy\right] \tag{26}$$

The total force balance along the main flow direction, derived in the Appendix, is also of interest:

$$\mathrm{Re}\frac{d}{dz}\int_{\kappa}^{1} U^2(r,z)r\,dr + \frac{d}{dz}\int_{\kappa}^{1} P(r,z)r\,dr + \frac{U(1,z)}{A}\exp(\varepsilon P(1,z)) + \kappa\frac{U(\kappa,z)}{A}\exp(\varepsilon P(\kappa,z)) = 0 \tag{27}$$

Equations (15), (25) and (26) constitute a system of non-linear integrodifferential equations which cannot be solved explicitly for the field variables $U$, $V$ and $P$. However, an approximate perturbation solution in terms of the small parameter $\varepsilon$ is derived below.

### 3.1. Perturbation solution in terms of ε

For all the dependent field variables, we assume regular perturbation expansions in terms of the non-negative parameter $\varepsilon$, which is assumed to be small. For $\varepsilon = 0$, i.e. in the case of Navier slip condition at the annular walls, the flow is fully developed, i.e. unidirectional. Thus, at the leading order of the perturbation $O(1)$, the radial velocity component is zero, the axial velocity component is a function of the radial distance only, and the pressure gradient along the pipe is constant. Once the $O(1)$ solution is known, one can proceed easily to the derivation of the $O(\varepsilon)$ and $O(\varepsilon^2)$ equations. Moreover, the solution procedure is greatly facilitated if the inner and outer slip velocities, i.e. $w(z) \equiv U(\kappa, z)$ and $W(z) \equiv U(1, z)$, are handled as separate unknown variables. Using, for brevity, the transformed axial coordinate $\hat{z} \equiv 1 - z$ and taking into account the above definitions and assumptions, the perturbation expansions of all the unknowns in terms of $\varepsilon$ are:

$$P = \underbrace{p_0 \hat{z}}_{P_{(0)}} + \varepsilon \underbrace{\left( p_{10}(r) + \hat{z} p_{11} + \hat{z}^2 p_{12} \right)}_{P_{(1)}} + \varepsilon^2 \underbrace{\left( p_{20}(r) + \hat{z} p_{21}(r) + \hat{z}^2 p_{22} + \hat{z}^3 p_{23} \right)}_{P_{(2)}} + O(\varepsilon^3) \tag{28}$$

$$V = \varepsilon \underbrace{V_1(r)}_{V_{(1)}} + \varepsilon^2 \underbrace{\left( V_{20}(r) + \hat{z} V_{21}(r) \right)}_{V_{(2)}} + O(\varepsilon^3) \tag{29}$$

$$U = \underbrace{U_0(r)}_{U_{(0)}} + \varepsilon \underbrace{\left( U_{10}(r) + \hat{z} U_{11}(r) \right)}_{U_{(1)}} + \varepsilon^2 \underbrace{\left( U_{20}(r) + \hat{z} U_{21}(r) + \hat{z}^2 U_{22}(r) \right)}_{U_{(2)}} + O(\varepsilon^3) \tag{30}$$

$$W = \underbrace{W_0}_{W_{(0)}} + \varepsilon \underbrace{\left( W_{10} + \hat{z} W_{11} \right)}_{W_{(1)}} + \varepsilon^2 \underbrace{\left( W_{20} + \hat{z} W_{21} + \hat{z}^2 W_{22} \right)}_{W_{(2)}} + O(\varepsilon^3) \tag{31}$$

$$w = \underbrace{w_0}_{w_{(0)}} + \varepsilon \underbrace{\left( w_{10} + \hat{z} w_{11} \right)}_{w_{(1)}} + \varepsilon^2 \underbrace{\left( w_{20} + \hat{z} w_{21} + \hat{z}^2 w_{22} \right)}_{w_{(2)}} + O(\varepsilon^3) \tag{32}$$

where the dependence on the radial and axial coordinates is explicit. The expansions in Eqs. (28)-(32) allow for a unique solution at zero-, first- and second-order in $\varepsilon$. More specifically, Eq. (25) is satisfied identically for any $\hat{z}$ by choosing the components of the radial velocity as follows:

$$V_1(r) = \frac{1}{r} \int_\kappa^r U_{11}(y) y \, dy, \quad V_{20}(r) = \frac{1}{r} \int_\kappa^r U_{21}(y) y \, dy, \quad V_{21}(r) = \frac{2}{r} \int_\kappa^r U_{22}(y) y \, dy \tag{33}$$

Similarly, Eq. (15) is satisfied identically by setting the pressure components as follows:

$$p_{10}(r) = a \left( V_1'(r) + \frac{V_1(r)}{r} - V_1'(1) \right) \tag{34}$$

$$p_{21}(r) = a \left( V_{21}'(r) + \frac{V_{21}(r)}{r} - V_{21}'(1) \right) \tag{35}$$

and

$$p_{20}(r) = a \left( V_{20}'(r) + \frac{V_{20}(r)}{r} - V_{20}'(1) \right) - a \operatorname{Re} \left( \frac{V_1^2(r)}{2} + \int_r^1 U_0(y) V_{21}(y) \, dy \right) \tag{36}$$

Note that the pressure expansion in Eq. (28), satisfies the reference pressure condition $P(r=1, z=1) = P(r=1, \hat{z}=0) = 0$. Thus, the r-momentum, Eq. (15), and the mass balance, Eq.(14), up to $O(\varepsilon^2)$ are eliminated, and the problem is reduced to the determination of the components of the velocity along the axial direction, $U_0, U_{10}, U_{11}, U_{20}, U_{21}$ and $U_{22}$ (all are unknown functions of the radial coordinate, $r$), of the pressure constants $p_0, p_{11}, p_{12}, p_{22}$ and $p_{23}$, and of the slip velocities $w_0, w_{01}, w_{11}, w_{20}, w_{21}$ and $w_{22}$, and $W_0, W_{01}, W_{11}, W_{20}, W_{21}$ and $W_{22}$. This task is accomplished by means of Eq. (26), the total force balance on the system, Eq. (27), and the integral constraint of the mass balance, Eq. (21). Note that the correctness of the perturbation solution for all the field variables has been confirmed with the 'Mathematica' symbolic software [27]. Before proceeding with the derivation of the solution, we introduce the auxiliary function $g = g(r)$, which appears frequently during the solution procedure:

$$g(r) = \int_{\kappa}^{r} \ln(y/r) y \, dy = \frac{1}{4}\left[\kappa^2 - r^2 + 2\kappa^2 \ln(r/\kappa)\right] = \frac{\kappa^2}{4}\left[1 - \left(\frac{r}{\kappa}\right)^2 + 2\ln\left(\frac{r}{\kappa}\right)\right] \tag{37}$$

Worth mentioning is that $\lim_{\kappa \to 0^+} g(r) = -r^2/4$.

### 3.2. Zero-order solution

From Eq. (26), we get the leading-order fluid velocity:

$$U_0(r) = w_0\left(1 + \frac{\kappa}{A}\ln\left(\frac{r}{\kappa}\right)\right) + p_0 \, g(r) \tag{38}$$

where $p_0$ is the zero-order pressure difference across the tube; see Eq. (28). This is trivially evaluated from Eq. (27):

$$p_0 = \frac{2(W_0 + \kappa w_0)}{A(1-\kappa^2)} \tag{39}$$

By means of

$$U_0(1) = W_0, \quad \int_{\kappa}^{1} U_0(r) r \, dr = 1 \tag{40}$$

one finds the leading-order slip velocities $W_0$ and $w_0$:

$$W_0 = \frac{4A\left[(2A-\kappa)(1-\kappa^2) - 2\kappa \ln(\kappa)\right]}{(1-\kappa)(1+\kappa)^2\left[A(\kappa-2)^2 - 3A + (4A^2-\kappa)(1-\kappa)\right] - \kappa\left[1-\kappa^4 + 4A(1+\kappa^3)\right]\ln(\kappa)} \tag{41}$$

and

$$w_0 = \frac{4A\left[(1+2A)(1-\kappa^2) + 2\kappa^2 \ln(\kappa)\right]}{(1-\kappa)(1+\kappa)^2\left[A(\kappa-2)^2 - 3A + (4A^2-\kappa)(1-\kappa)\right] - \kappa\left[1-\kappa^4 + 4A(1+\kappa^3)\right]\ln(\kappa)} \tag{42}$$

Notice that $w_0$ and $W_0$ range from 0 ($A=0$; no slip) to $2/(1-\kappa^2)$ ($A \to \infty$; full slip in which case the velocity profile is uniform and $w_0 = W_0$). Taking into account the different scalings used in [20, 22], the present zero-order solution is equivalent to the expressions provided in these two references for the case of standard (pressure-independent) Navier slip.

### 3.3. First-order solution

From Eq. (26) at $O(\varepsilon)$, we find:

$$U_{10}(r) = w_{10}\left[1+\frac{\kappa}{A}\ln\left(\frac{r}{\kappa}\right)\right] + p_{11}g(r) + \mathrm{Re}\int_{\kappa}^{r}U_0(y)\left[V_1(y)+2U_{11}(y)\ln\left(\frac{y}{r}\right)y\right]dy \tag{43}$$

and

$$U_{11}(r) = w_{11}\left[1+\frac{\kappa}{A}\ln\left(\frac{r}{\kappa}\right)\right] + 2p_{12}g(r) + \frac{\kappa}{A}w_0 p_0 \ln\left(\frac{r}{\kappa}\right) \tag{44}$$

From the total momentum balance along the z-direction at $O(\varepsilon)$ and collecting the terms with respect to $\hat{z}$, we get the pressure components:

$$p_{11} = \frac{2(W_{10}+\kappa w_{10})}{A(1-\kappa^2)} - \frac{4\mathrm{Re}}{1-\kappa^2}\int_{\kappa}^{1}U_0(r)U_{11}(r)rdr \tag{45}$$

and

$$p_{12} = \frac{W_{11}+\kappa w_{11}}{A(1-\kappa^2)} + \frac{p_0^2}{2} \tag{46}$$

The integral in Eq. (45) is calculated by substituting the solution for $U_{11}$:

$$\int_{\kappa}^{1}U_0(r)U_{11}(r)rdr = w_{11} + \frac{\kappa}{A}(w_0 p_0 + w_{11})\underbrace{\int_{\kappa}^{1}U_0(r)\ln\left(\frac{r}{\kappa}\right)rdr}_{I_a(U_0)} + 2p_{12}\underbrace{\int_{\kappa}^{1}U_0(r)g(r)rdr}_{I_b(U_0)} \tag{47}$$

where $I_a(U_0)$ and $I_b(U_0)$ are known functions of $\kappa$ and $A$. Substituting the above integral in Eq. (45) for $p_{11}$ and simplifying, we get:

$$p_{11} = \frac{2(W_{10}+\kappa w_{10})}{A(1-\kappa^2)} - \frac{4\mathrm{Re}}{1-\kappa^2}\left\{w_{11} + \frac{\kappa}{A}(w_0 p_0 + w_{11})I_a(U_0) + \left[\frac{2(W_{11}+\kappa w_{11})}{A(1-\kappa^2)} + p_0^2\right]I_b(U_0)\right\} \tag{48}$$

and

$$p_{12} = \frac{W_{11}+\kappa w_{11}}{A(1-\kappa^2)} + \frac{p_0^2}{2} \tag{49}$$

The first-order solution is completed by solving the equations

$$U_{1j}(1) = W_{1j}, \quad \int_{\kappa}^{1}U_{1j}(r)rdr = 0 \tag{50}$$

for $j=1$ to determine $W_{11}$ and $w_{11}$, and likewise for $j=0$ to determine $W_{10}$ and $w_{10}$.

### 3.4. Second-order solution

The solution procedure at second order is similar, but the resulting expressions are too lengthy to present here due to the appearance of quadratic terms with respect to $\hat{z}$. The flow variables have been determined analytically, and their solutions, along with the solutions at zero and first orders in ε, are provided in the supplementary material [28] as 'Mathematica' notebooks.

## 4. Results and Discussion

All the results presented in this Section were obtained assuming $\alpha = 0.01$. Before proceeding with the discussion of the main results, we first investigate the ranges of the dimensionless parameters for which the perturbation solution is valid. Since the exponential function is always positive, the same should be held for its series expansion. In particular, we demand that the perturbed exponential term that appears in the slip law is positive up to second-order:

$$\exp(-\varepsilon P) \approx 1 - \varepsilon P_{(0)} + \varepsilon^2 \left( \frac{P_{(0)}^2}{2} - P_{(1)} \right) > 0 \qquad (51)$$

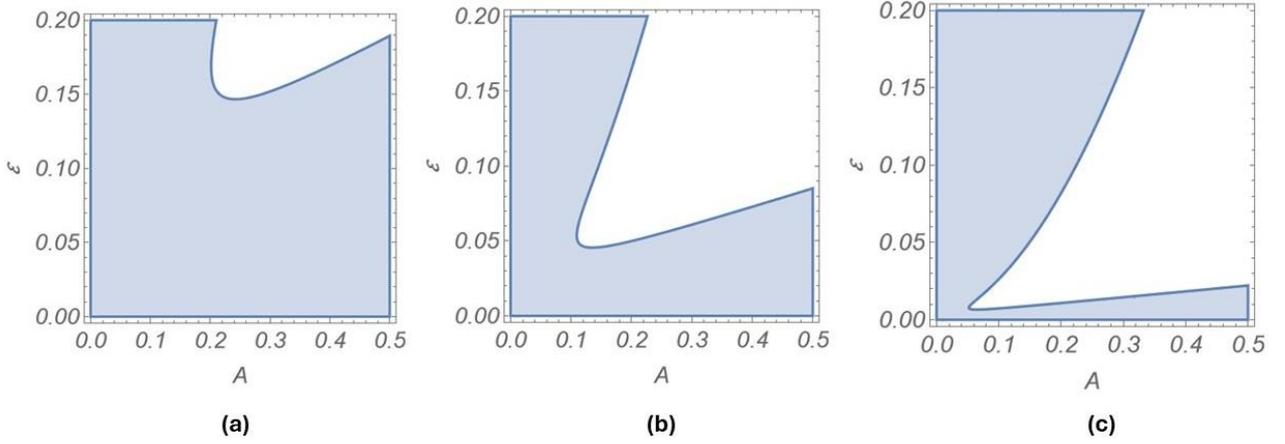

**Figure 2.** The admissible range of parameters ε and A (shaded area) for which exp(-ε P(r=1,z)) > 0 for any value of the z-coordinate; parameters are α=0.01 and Re=1. (a) κ=0.1; (b) κ=0.4; (c) κ=0.7.

In Fig. 2, the grey (shaded) area represents the range of parameters on the (ε,A) plane for which $\exp(-\varepsilon P(r=1,z))$ is positive for any value of the z-coordinate. The results are shown for κ=0.1, 0.4, and 0.7, while the other parameters are $\mathrm{Re}=1$ and $\alpha = 0.01$. As expected, the positivity of the exponential function is violated when both ε and A are large. It is also observed that an increase in the radii ratio shrinks the admissible range of parameters, and as $\kappa \rightarrow 1^-$, $\varepsilon \rightarrow 0^+$. Notably, due to the complex dependence of the analytical solution on A, the curve that separates the admissible from the non-admissible region is difficult to describe.

As mentioned in Section 3, the zero-order solution $U_0(r)$ corresponds to the standard Poiseuille flow solution in the presence of the classic Navier slip at the walls, where the axial velocity varies only with r, the radial velocity component vanishes, and the pressure linearly with z. When wall slip is pressure dependent, the axial velocity profile deviates from the classic Poiseuille solution and varies with both z and r, $U = U(r,z)$. Consequently, a radial velocity component is generated too due to the continuity equations, i.e. $V = V(r,z)$. In Fig. 3 we plotted the deviation $\delta u_z = U(r,z) - U_0(r)$ at various axial positions for $\text{Re} = 0.1$ and $\varepsilon$ close to the maximum admissible value of $\varepsilon$ (by means of the exponential function criterion) for $\kappa = 0.1$ and $\kappa = 0.5$.

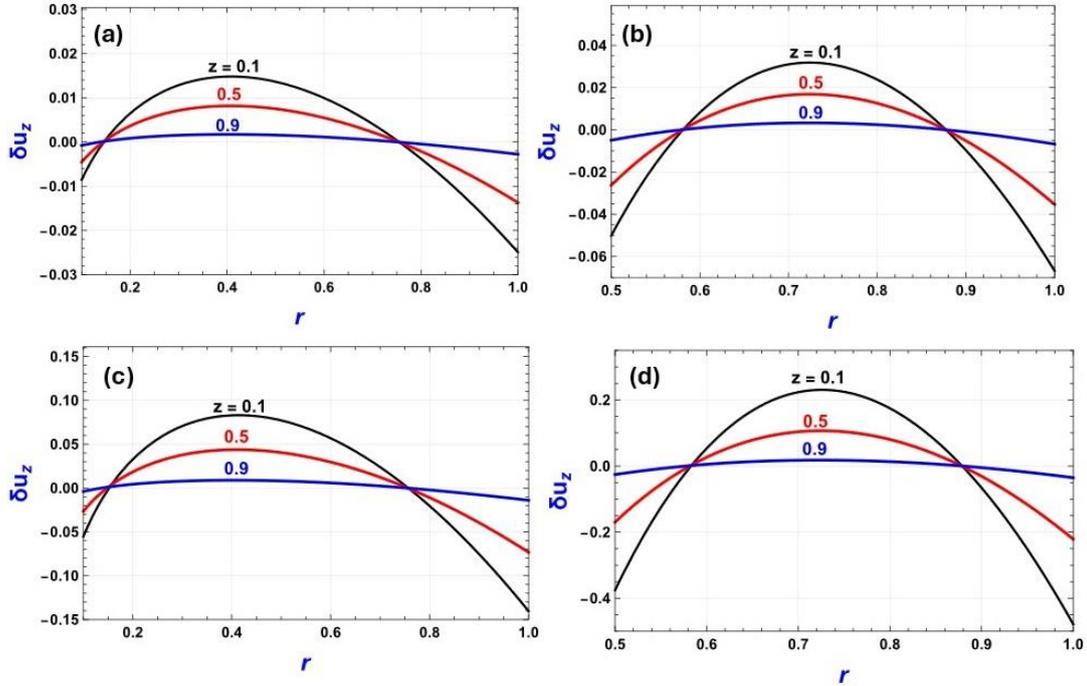

**Figure 3.** Deviations from the zero-order profile, $\delta u_z(r,z) = U(r,z) - U_0(r)$, at selected distances downstream when $A=0.5$, $\alpha=0.01$ and Re=1: (a) $\kappa=0.1$, $\varepsilon=0.01$; (b) $\kappa=0.5$, $\varepsilon=0.01$; (c) $\kappa=0.1$, $\varepsilon=0.05$; (d) $\kappa=0.5$, $\varepsilon=0.05$.

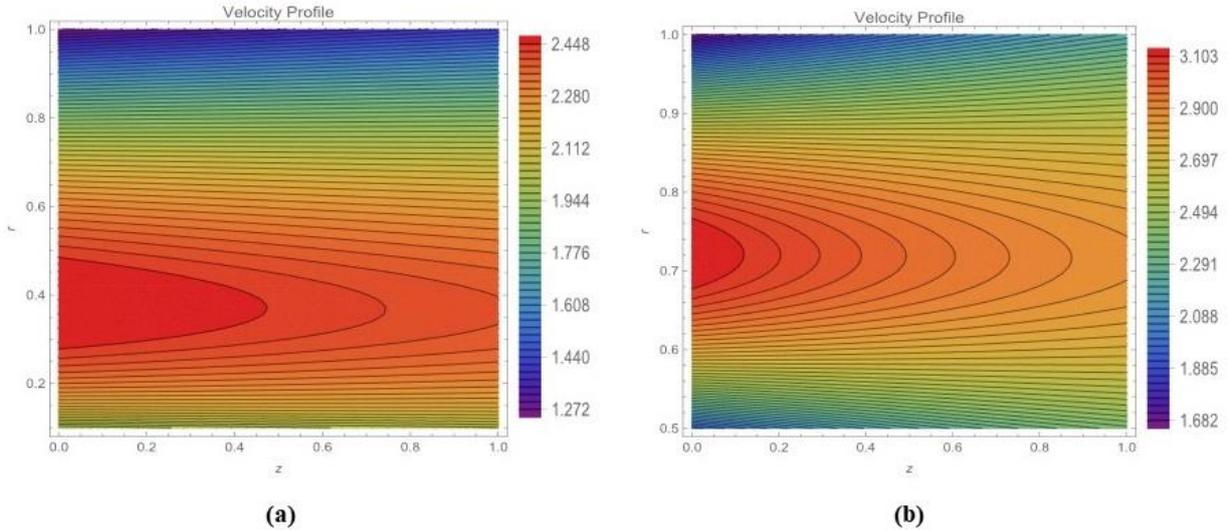

**Figure 4.** Axial velocity contours for Re=1, $A=0.5$, and $\varepsilon=0.05$; (a) $\kappa=0.1$, (b) $\kappa=0.5$.

In Fig. 4, the contours of the axial velocity $U(r,z)$ for $\text{Re}=1$, $A=0.5$ and $\varepsilon=0.05$ are provided for two values of the radii ratio, $\kappa=0.1$ and 0.5. As expected, the velocity profile tends to become symmetric as $\kappa$ is increased. We observe that for bigger radii ratios the velocity profile tends to become symmetric. Theoretically, as the gap size becomes narrower ($\kappa \to 1$), the flow approaches the corresponding channel flow. Upon closer examination, we observe that the velocity near the walls increases upstream while the opposite holds at the radial position where the velocity reaches its maximum. The latter observations regarding the wall velocities are also verified in Fig. 5. Besides, increasing the slip decay coefficient $\varepsilon$ reduces the wall velocities downstream. As expected, when $\kappa = 0.5$ the difference between the two wall velocities decreases, while as $\kappa \to 1$ the two slip velocities tend to become equal.

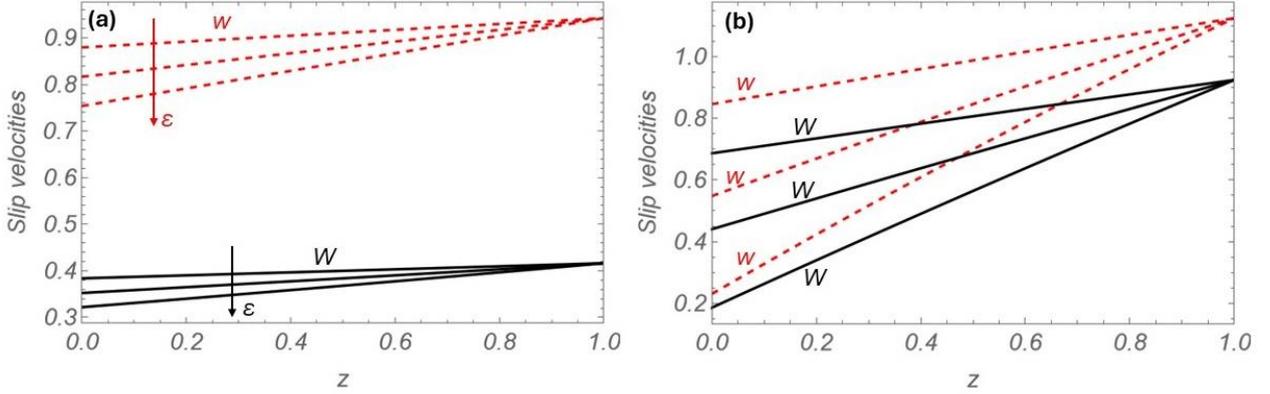

**Figure 5.** Inner slip velocity (dashed red lines), w(z)=U(κ,z), and outer slip velocity (solid black lines), W(z)=U(1,z), as functions of the axial direction from the inlet, z, for ε=0.005, 0.010, and 0.015; Re=1 and A=0.05. (a) κ=0.1; (b) κ=0.5.

The shear stress,

$$\tau_{rz} = \frac{\partial U}{\partial r} + a\frac{\partial V}{\partial z} \approx \frac{\partial U_{(0)}}{\partial r} + \varepsilon \frac{\partial U_{(1)}}{\partial r} + \varepsilon^2 \frac{\partial U_{(2)}}{\partial r} - \varepsilon^2 a V_{21}(r) \tag{52}$$

is of interest in internal and confined flows (particularly along the walls), since it is directly related to the average pressure drop required to drive the flow. Equation (52) is derived from Eqs. (29) and (30) up to $O(\varepsilon^2)$. Figs. 6 and 7 depict how the shear stress varies in the flow domain when $\kappa = 0.1$ and 0.5, respectively, for $\mathrm{Re} = 0.1$, $A = 0.5$, and $\varepsilon = 0.01$ and 0.05. In the case of $\varepsilon = 0.01$ (Fig. 6) one observes that the shear stress grows in magnitude downstream, and this effect is enhanced with increasing $\varepsilon$. In addition, the shear stress reaches its absolute maximum value at the inner wall whilst it changes sharply with the radial distance *r*. The latter indicates that the maximum velocity is located close to the inner wall. Another observation is that the radial position at which the velocity takes its maximum changes across the annular tube. Similar observations hold when $\kappa = 0.5$ (Fig. 7), but in this case the shear stress distribution is less asymmetric.

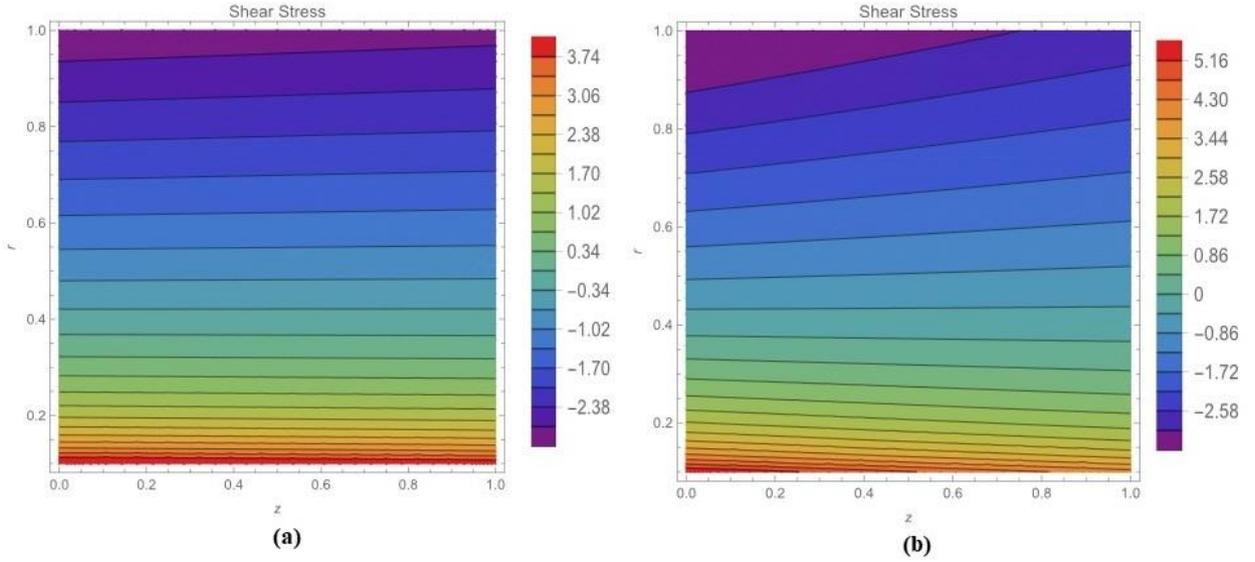

Figure 6. Shear stress contours for Re=0.1, A=0.5, and κ=0.1. (a) ε=0.01; (b) ε=0.05.

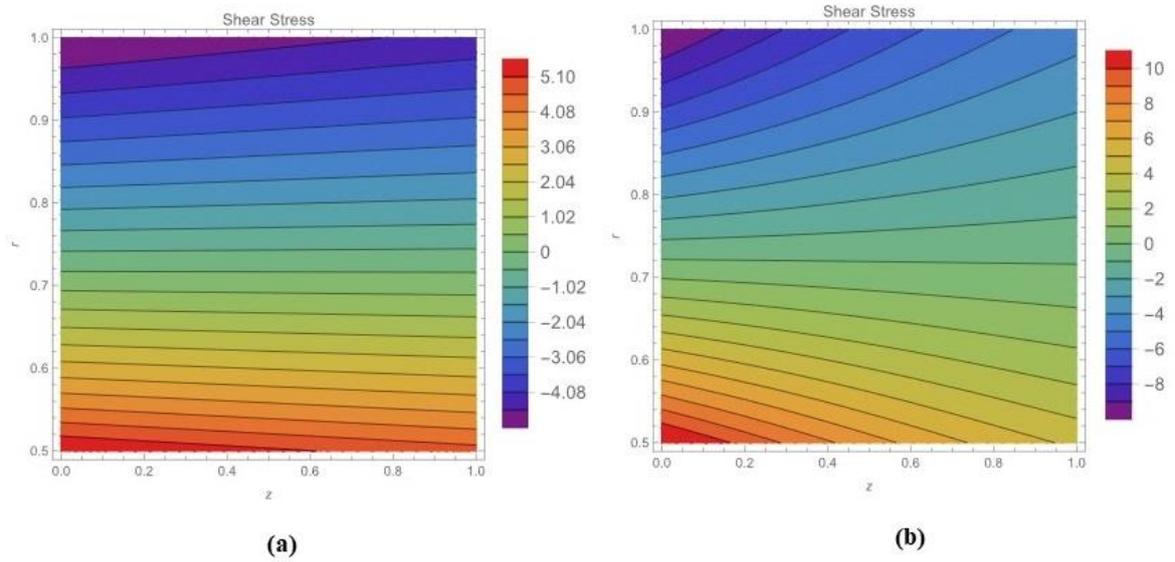

Figure 7. Shear stress contours for Re=0.1, A=0.5, and κ=0.5. (a) ε=0.01; (b) ε=0.05.

### 4.1. Average pressure-drop

The average pressure-drop along the annulus, $\Delta\langle p \rangle$, is a critical parameter for designing and optimizing fluid flow systems, such as pipelines, ventilation systems, and chemical reactors, ensuring efficient and safe operation. Given that the pressure varies with both *r* and *z*, we consider the difference of the average pressures between the inlet and the outlet planes:

$$\Delta\langle p \rangle \equiv 2\int_{\kappa}^{1}\left[P(r,0)-P(r,1)\right]r\,dr \tag{53}$$

where in general $\Delta f := f(r,z=0) - f(r,z=1)$ for any function $f = f(r,z)$. An alternative derivation of $\Delta\langle p \rangle$ follows from the total force balance on the system, which is obtained by integrating Eq. (27) with respect to the axial coordinate from the inlet to the outlet cross-sections of the annulus:

$$\Delta \langle p \rangle = -2 \operatorname{Re} \Delta \left( \int_{\kappa}^{1} U^2(r,z) r dr \right) + \frac{2}{A} \int_{0}^{1} (W(z) e^{\varepsilon P(1,z)} + \kappa w(z) e^{\varepsilon P(\kappa,z)}) dz \qquad (54)$$

Equation (54) shows that the average pressure drop results from two contributions: the difference in kinetic energy between the output and inlet cross-sections of the tube, and the kinetic energy transferred along the walls of the annulus due to slip. While the sign of the former is uncertain, the sign of the latter is clearly positive (given that the slip velocities are positive). We have confirmed this by substituting the perturbation expansions into Eqs. (53) and (54), suitably expanding all quantities, and collecting like terms of $\varepsilon$, we obtain the same result for the average pressure-drop, clearly indicating the correctness and consistency of our methods. Specifically, we find:

$$\Delta \langle p \rangle \approx \Delta P_{(0)} + \varepsilon \Delta P_{(1)} + \varepsilon^2 \Delta P_{(2)} \qquad (55)$$

where

$$\Delta P_{(0)} = \frac{2}{A}(W_0 + \kappa w_0) \qquad (56)$$

and

$$\Delta P_{(1)} = -2\operatorname{Re}\left( \int_{\kappa}^{1} U_0(r) U_{11}(r) r\, dr \right) + \frac{2}{A}\left( W_{10} + \kappa w_{10} + \frac{W_{11} + \kappa w_{11}}{2} \right) + \frac{\Delta P_{(0)}^2}{2(1-\kappa^2)} \qquad (57)$$

while $\Delta P_{(2)}$ is too long to be printed here.

To increase the accuracy of Eq. (55), we utilize the Padé [1/1] diagonal formula [29,30,31]:

$$\Delta \langle p \rangle_{acc} = \Delta P_{(0)} + \frac{\varepsilon \Delta P_{(1)}^2}{\Delta P_{(1)} - \varepsilon \Delta P_{(2)}} \qquad (58)$$

Equation (58) represents the simplest and lowest-order non-linear approximant of a function when only the first three terms in its Taylor series are available. It is also known to increase the accuracy and extend the range of validity of the original truncated series up to $O(\varepsilon^2)$, except in cases where $\Delta P_{(1)}$ and $\Delta P_{(2)}$ have the same sign and the critical value $\varepsilon_c = \Delta P_{(1)} / \Delta P_{(2)}$ is not within the interval of interest for $\varepsilon$ [30,31,32]. In this case, $\varepsilon_c$ is usually a spurious singular point of the solution and Eq. (58) cannot be used in the vicinity of $\varepsilon_c$. Note, however, that within the range of parameters investigated here, no singular points for $\varepsilon$ were encountered. Finally, it is important to mention that it is not straightforward to determine how much better Eq. (58) is compared to the original truncated series; to assess this, higher-order terms in the series involving $\varepsilon$ are needed. Below we present the results by using the normalizations $\Delta\langle p\rangle/\Delta P_{(0)}$ and $\Delta_{acc}\langle p\rangle/\Delta P_{(0)}$ which facilitates comparisons and reveals the net effect of the small parameter $\varepsilon$ on the pressure drop. Based on our experience in a variety of fluid mechanics problems solved with perturbation methods [30,31,32,33], we believe that the results for $\Delta_{acc}\langle p\rangle/\Delta P_{(0)}$ are more accurate than $\Delta\langle p\rangle/\Delta P_{(0)}$; however, we present both for completeness and comparisons.

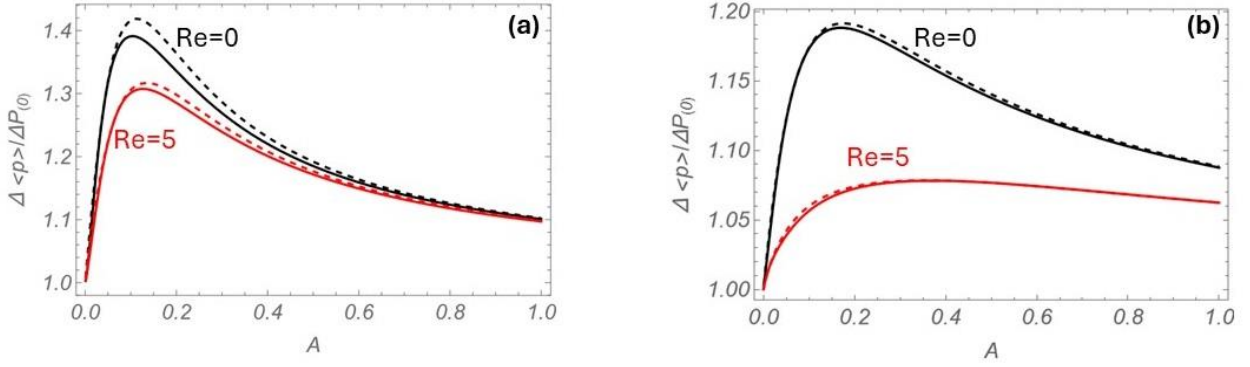

**Figure 8.** Reduced average pressure drop as function of the slip number for Re=0 and 5. Solid and dashed lines are the results from Eqs. (55) and (58), respectively. (a) κ=0.5 and ε=0.01; (b) κ=0.1 and ε=0.05.

The effect of the slip number, $A$, on the reduced average pressure drop, $\Delta\langle p\rangle/\Delta P_{(0)}$, is illustrated in Fig. 8a for $\kappa=0.1$ and $\varepsilon=0.05$, and in Fig. 8b for $\kappa=0.5$ and $\varepsilon=0.01$. The Reynolds number is $\mathrm{Re}=0$ and 5. For very small values of the slip number, the reduced pressure drop increases sharply. This increase is due to the variation of the slip velocities with the distance from the inlet, as shown earlier in Fig. 4 for $\varepsilon>0$. Indeed, in this case, steeper gradients are generated in both spatial directions near the wall compared to the constant slip case (i.e., for $\varepsilon=0$), resulting in larger pressure differences required to drive the flow. However, as the slip number increases further, i.e., as slip becomes stronger, the wall velocities increase in magnitude, but their spatial gradients decrease, thereby reducing the required pressure drop. Thus, a maximum value of $\Delta\langle p\rangle/\Delta P_{(0)}$ at moderate slip is predicted. Regarding the effect of inertia, the results reveal a slight reduction in the pressure drop for small and moderate slip numbers, but for strong slip, the effect of inertia is minor, as the velocity profile across the annulus becomes more plug-like without steep gradients. Last, the good agreement of $\Delta\langle p\rangle/\Delta P_{(0)}$ with $\Delta_{acc}\langle p\rangle/\Delta P_{(0)}$ indicates the convergence of the results for the parameters values shown in the figure.

Last, In Fig. 9, we plot the reduced average pressure drop as a function of the Reynolds number for various slip numbers and two gap sizes (κ=0.1 and κ=0.5). The most important information extracted from these results is that the pressure drop remains constant with the Reynolds number until the Reynolds number becomes sufficiently large, after which the pressure drop decreases monotonically. The larger the initial pressure drop, the sooner it begins to decrease. However, we emphasize that accurately determining the variation of the pressure drop at high Reynolds numbers requires higher-order terms in the perturbation series, more accurate Padé or other non-linear approximants, or numerical methods.

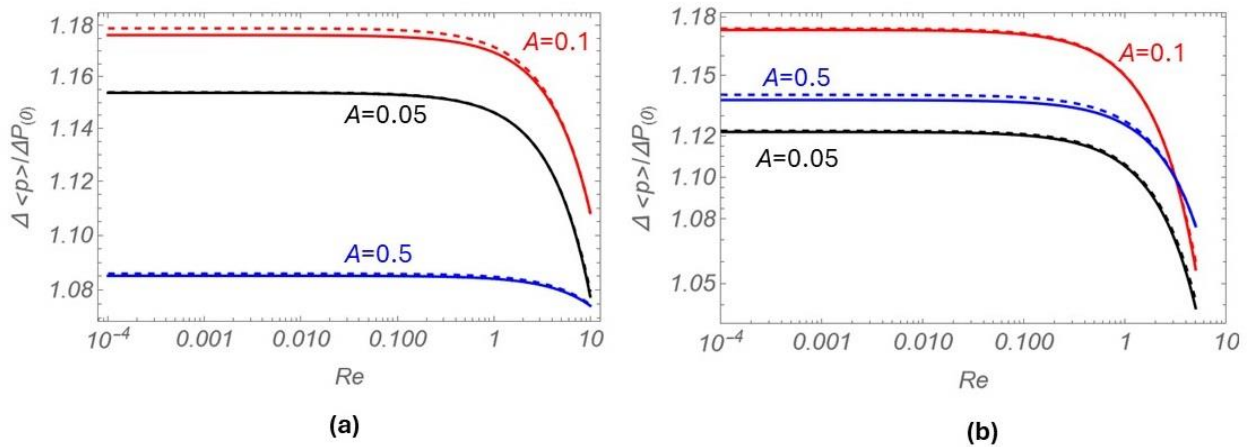

Figure 9. Reduced average pressure drop as a function of the Reynolds number for $A$=0.05, 0.1 and 0.5. Solid and dashed lines are the predictions of Eqs. (55) and (58), respectively. (a) $\kappa$=0.5 and $\varepsilon$=0.01; (b) $\kappa$=0.1 and $\varepsilon$=0.05.

## 5. Conclusions

We have reported the results of the perturbation solution of laminar, incompressible and isothermal annular Newtonian Poiseuille flow with wall slip assuming an exponential dependence of the slip coefficient on the pressure. The original dimensionless governing equations have been converted in non-linear integrodifferential form before a perturbation scheme up to second order was applied. The pressure drop has been obtained by utilizing an analytical formula and have been optimized by applying a Padé diagonal non-linear transformation. The solution shows that the effect of pressure on slip gradually diminishes as the constant slip coefficient increases which implies that the slip velocities tend to be constant throughout the annular tube. As the radii ratio increases the effects of pressure on slip become more pronounced. For small annular gaps the flow becomes more symmetrical tending asymptotically to plane Poiseuille flow.

    The derived approximate analytical solution can be useful for validating numerical algorithms for annular flows with pressure-dependent wall slip. A natural extension of this work is to take into account the pressure-dependence of the viscosity and the other rheological parameters in generalized Newtonian flows [34,35,36].

**Data availability**
The data that support the findings of this study are available in [28].

## Appendix A: Solution of the momentum balance (Eq. (16))

First, we note that for any continuous function $f = f(r,z)$, where $\kappa \leq r \leq 1$ and $0 \leq z \leq 1$, the following identity holds:

$$\int_\kappa^r \frac{1}{s}\left(\int_\kappa^s f(y,z) y\, dy\right) ds = -\int_\kappa^r f(y,z) \ln(y/r) y\, dy \tag{A1}$$

Then, we reformulate Eq. (15) in conservative form:

$$\text{Re}\left(\frac{\partial}{\partial r}(rVU) + \frac{\partial}{\partial z}(rU^2)\right) = -\frac{\partial}{\partial z}(rP) + \frac{\partial}{\partial r}\left(r\frac{\partial U}{\partial r}\right) + a\frac{\partial^2}{\partial z^2}(rU) \tag{A2}$$

Integrating Eq. (A2) with respect to $r$, from $\kappa$ to $r$, and using the no-penetration condition at the inner wall give:

$$\text{Re}\left(rVU + \frac{\partial}{\partial z}\int_\kappa^r U^2(y,z) y\, dy\right) = -\frac{\partial}{\partial z}\int_\kappa^r P(y,z) y\, dy + r\frac{\partial U}{\partial r} - \kappa\frac{\partial U}{\partial r}\bigg|_{r=\kappa} + a\frac{\partial^2}{\partial z^2}\int_\kappa^r U(y,z) y\, dy \tag{A3}$$

By means of the slip condition on the inner wall (Eq. (19)) one gets:

$$\text{Re}\left(VU + \frac{1}{r}\frac{\partial}{\partial z}\int_\kappa^r U^2(y,z) y\, dy\right) = -\frac{1}{r}\frac{\partial}{\partial z}\int_\kappa^r P(y,z) y\, dy + \frac{\partial U}{\partial r} - e^{\varepsilon P(\kappa,z)}\frac{\kappa}{Ar}U(\kappa,z) + \frac{a}{r}\frac{\partial^2}{\partial z^2}\int_\kappa^r U(y,z) y\, dy \tag{A4}$$

Integrating Eq. (A4) with respect to $r$, from $\kappa$ to $r$, gives:

$$U(r,z) = U(\kappa,z)\left[1 + \frac{\kappa}{A}e^{\varepsilon P(\kappa,z)}\ln\left(\frac{r}{\kappa}\right)\right] + \text{Re}\left(\int_\kappa^r I(s,z)\, ds\right) + \frac{\partial}{\partial z}\int_\kappa^r \frac{1}{s}\left(\int_\kappa^s p(y,z) y\, dy\right) ds - a\frac{\partial^2}{\partial z^2}\int_\kappa^r \frac{1}{s}\left(\int_\kappa^s U(y,z) y\, dy\right) ds \tag{A5}$$

where

$$I(r,z) := U(r,z)V(r,z) + \frac{1}{r}\frac{\partial}{\partial z}\int_\kappa^r U^2(y,z) y\, dy \tag{A6}$$

However,

$$\int_\kappa^r I(s,z)\, ds = \int_\kappa^r U(s,z)V(s,z)\, ds + \frac{\partial}{\partial z}\int_\kappa^r \frac{1}{s}\left(\int_\kappa^s U^2(y,z) y\, dy\right) ds = \int_\kappa^r U(y,z)V(y,z)\, dy - \frac{\partial}{\partial z}\int_\kappa^r \ln\left(\frac{y}{r}\right) U^2(y,z) y\, dy \tag{A7}$$

Substituting Eq. (A7) into Eq. (A5) gives:

$$U(r,z) = U(\kappa,z)\left[1 + \frac{\kappa}{A}e^{\varepsilon P(\kappa,z)}\ln\left(\frac{r}{\kappa}\right)\right] + \text{Re}\left(\int_\kappa^r U(y,z)V(y,z)\, dy\right) - \frac{\partial}{\partial z}\left[\int_\kappa^r \ln\left(\frac{y}{r}\right)\left(\text{Re}\, U^2(y,z) + p(y,z)\right) y\, dy\right] + a\frac{\partial^2}{\partial z^2}\left[\int_\kappa^r \ln\left(\frac{y}{r}\right) U(y,z) y\, dy\right] \tag{A8}$$

Finally, evaluating Eq. (A4) at $r=1$ and using the corresponding slip condition, Eq. (18), and the total mass balance, Eq. (21), gives Eq. (27). The derivation is completed by imposing the reference pressure, Eq. (22).